# AUTOMATIC CLASSIFICATION OF X-RATED VIDEOS USING OBSCENE SOUND ANALYSIS BASED ON A REPEATED CURVE-LIKE SPECTRUM FEATURE

JaeDeok Lim[1], ByeongCheol Choi[1], SeungWan Han[1] and ChoelHoon Lee[2]

[1] Knowledge-based Information Security Research Division, ETRI, Daejeon, KOREA
{jdscol92, corea, hansw}@etri.re.kr
[2] Department of Computer Engineering, Chung-Nam National University, Daejeon, KOREA
clee@cnu.ac.kr

## ABSTRACT

*This paper addresses the automatic classification of X-rated videos by analyzing its obscene sounds. In this paper, obscene sounds refer to audio signals generated from sexual moans and screams during sexual scenes. By analyzing various sound samples, we determined the distinguishable characteristics of obscene sounds and propose a repeated curve-like spectrum feature that represents the characteristics of such sounds. We constructed 6,269 audio clips to evaluate the proposed feature, and separately constructed 1,200 X-rated and general videos for classification. The proposed feature has an F1-score, precision, and recall rate of 96.6%, 98.2%, and 95.2%, respectively, for the original dataset, and 92.6%, 97.6%, and 88.0% for a noisy dataset of 5dB SNR. And, in classifying videos, the feature has more than a 90% F1-score, 97% precision, and an 84% recall rate. From the measured performance, X-rated videos can be classified with only the audio features and the repeated curve-like spectrum feature is suitable to detect obscene sounds.*

## 1. INTRODUCTION

The rapid development of multimedia technologies and advances in internet infrastructure have allowed general users to easily create, edit, and post their own content, and it is even easier to access any Internet content if so desired [1]. However, this has also led to a harmful side effect, which is the creation and distribution of uncontrolled X-rated videos. This is a particularly serious situation for pornographic videos [2]. Pornographic content makes up more than 70% of X-rated videos. In this paper, X-rated videos refers to pornographic or similar material, and obscene sounds are audio signals generated from sexual moans and screams during various sexual scenes. Recent sexual crime-related articles and reports show that X-rated videos can have impact on sexual crimes, both directly and indirectly. This is even more serious for youth and teenagers. Thus, there is a strong demand for more efficient technologies that can detect and classify X-rated videos for a clean Internet environment.

Most of studies on the X-rated video classification and detection have focused on image or video processing technologies. The Wavelet Image Pornography Elimination(WIPE) that is a multi-step screening method uses a manually-specified color histogram model as a pre-filter in an analysis pipeline [3]. Some studies uses skin color model as image-based features [4][7]. Especially the scheme of [4] has been incorporated and deployed into Google's adult-content filtering infrastructure for image safe-search [5]. For detecting pornographic video contents, it is used to combine image features of key frame with motion features by periodicity detection based on auto-correlation [8]. But such vision-based classification methods make it computationally expensive to carry out visual data processing and a large amount of storage is required for model parameters. Furthermore, images can easily be too poor of quality to be used for classification due to common degradations such as poor lightings and visual obstructions [6]. Also the classification with skin color is ineffective because general images can easily be classified as pornographic image if they contain a large amount of skin color, such as close up of face. Therefore, the image or video-based X-rated video classification systems may produce high false positive rates [6][9].

Unlike a lot of studies related to X-rated videos classification based on the aforementioned vision-based approaches, there have been relatively few studies based on audio-based approaches. The advantages of audio-based approaches for content classification include the need for fewer computational resources and less feature space compared to vision-based approaches [10]. Thus, audio-based approaches can compensate for the drawbacks of a vision-based classification system with much fewer computations and less memory.

In this paper, we propose the repeated curve-like spectrum feature (RCSF) as an audio feature for classifying obscene sounds and X-rated videos. For a reasonable evaluation of the proposed feature, we constructed two types of datasets: One is for generating a learning model and testing the performance of classifying obscene sounds and was constructed using 6,269 audio clips of 10s in length. The other is for classifying videos using the learning model generated from audio clips based on the RCSF feature and was constructed using 600 X-rated and 600 general video files. The proposed feature is compared to other well-known low-level spectral features and mel-frequency cepstrum coefficients (MFCCs) and their family features. A support vector machine (SVM) classifier with a radial basis kernel function (RBF) is used to learn the features extracted from audio clips for training and to classify them for testing.

The remainder of this paper is organized as follows. In section 2, we review some related works. In section 3, we analyse the distinguishable characteristics of obscene sound samples from non-obscene ones. The RCSF feature and feature vector construction are reviewed in section 4. The SVM classifier and its application in our experiments are explained in section 5. Construction of two types of datasets is described in detail in section 6. The experimental results are presented in Section 7. Finally, some concluding remarks are provided in section 8.

## 2. RELATED WORKS

Unlike the many studies related to X-rated videos classification based on vision-based approaches, there are relatively few studies based on audio-based approaches. An example of studies related to audio feature based classification and detection of X-rated videos uses the fact that sexual audio signals are periodically generated by specific activities of the actors [11]. This periodic pattern of sound can be estimated based on the audio periodicity used as an audio feature in detecting sexual scenes. In [11], audio periodicity is computed through auto-correlation of the signal energy, as periodicity of a signal is usually analyzed through auto-correlation. However, since the feature of periodicity depends only on the periodicity of sound, it is obvious that the performance is poor when periodic sounds do not occur in sexual scenes or when a non-sexual scene is composed of periodic sounds such as during ping-pong or tennis matches. Also, it is not a proper audio feature to classify obscene sound in real fields, as most obscene sounds within various videos are composed of background music, some recording noises, other environment sounds, and so on, as well as main sounds. In fact, the audio periodicity cannot discriminate obscene sounds from non-obscene sounds in our dataset, which will be explained in section 6.

The correlation coefficient of an audio signal spectrum is used as an audio-based feature to detect obscene sounds [12]. In [12], the correlation coefficient is computed using the spectrum of the target sound to be tested and the spectrum of tens of standard obscene sounds one by one. Standard obscene sounds that are unrelated with each other are selected. If one of the computed correlation coefficients is larger than the threshold value (50% in [12]), the target sound is classified as an obscene sound. Through the correlation coefficient feature, it is difficult to detect obscene sounds within various contents for the same reasons that are explained for periodicity feature, and the feature may have strong dependency with standard signals. If the number of standard signals is increased for expanding the coverage of obscene sounds, the computing overhead is increased with an increased number of signals, and it is still difficult to improve the detection performance. In our test on 100 standard obscene sounds, 100 testing obscene sounds, and 100 testing non-obscene sounds selected from our audio clip database, which will be explained in section 6, the average classification performances for obscene sounds with 0.5, 0.6, and 0.7 threshold values are 62.86%, 33.93%, and 9.2% respectively. Also, the average classification performances for non-obscene sounds under 0.5, 0.6, and 0.7 threshold values are 37.24%, 16.98%, and 5.39%, respectively. The results show that the correlation coefficient of an audio signal spectrum is difficult to use as an audio feature to detect a variety of obscene sounds.

Our earlier study had tried to classify obscene sounds using various MFCC-based feature sets [13]. The most accurate rate was about 90% for a 15-order MFCC feature set. However, the dataset used in [13] is not large enough, and the feature set is limited to only MFCC. This is needed to increase the size of the dataset and conduct a test using more varied feature sets.

## 3. ANALYSIS OF OBSCENE SOUND CHARACTERISTICS

By analyzing many obscene sounds, we can find that most obscene sounds within sexual scenes are composed of human-based sounds; activity-based sounds from human bodies such as deep kisses, oral activities, and so on; and background sounds, which can be treated as environmental noise such as music or bed creaking. We focus on human-based sounds since most obscene sounds within sexual scenes consist of mostly human-based sounds. Obscene human-based sounds can be divided into certain categories based on sound sources and sound types. Sound sources are female (FM), male (MA), and both female and male (BO), and sound types are a hard sexual scream (SS) and soft sexual moan (SM). From this division, we can categorize human-based obscene sounds with the combinations of sound sources and sound types, that is, FMSS, FMSM, MASS, MASM, BOSS, and BOSM. For example, FMSS means a hard sexual scream sound generated from mostly a female. The dataset of audio clips for obscene class will be constructed according to these categories in section 6.

To define the characteristics of obscene sounds that are more generally distinguished from non-obscene sounds, we have analyzed the spectrogram patterns of various types of sounds. Figures and 2 show the typical spectrogram patterns of obscene and non-obscene sounds.

The International Journal of Multimedia & Its Applications (IJMA) Vol.3, No.4, November 2011

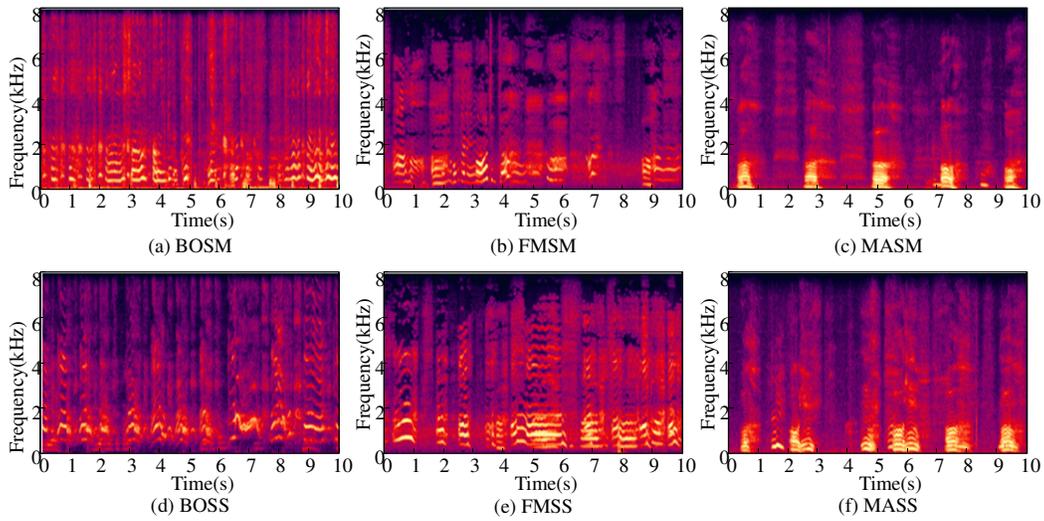

Figure 1. Spectrograms of obscene sounds

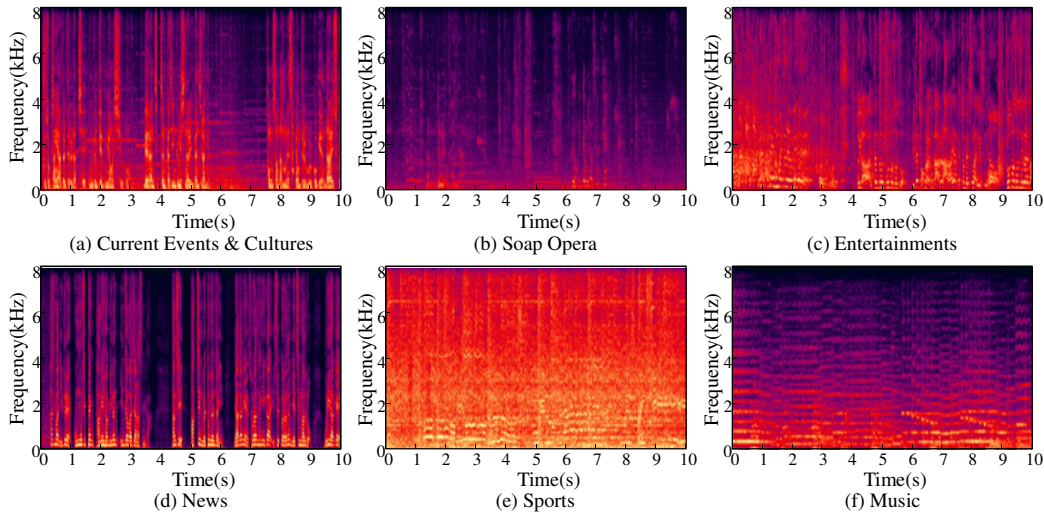

Figure 2. Spectrograms of non-obscene sounds

The most distinguished characteristic is that obscene sounds have a repeated curve-like pattern in all of the six categories mentioned above. Even though curve-like patterns sometimes appear in non-obscene sounds, they are not clear and are not repeated. However, in obscene sounds, curve-like patterns of about 500ms in length are shown to clearly repeat. This means that the frequency spectrum of obscene sounds has a repeatedly temporal variation. Also, while a general speech signal has a low pitch within the range of 100-300Hz, most obscene sounds have a relatively high pitch, typically about 500Hz, and often include heavy breathing that leads to a high energy noise property at more than 4 kHz. We focus on modeling the repeated temporal variation of the frequency spectrum, and use it as a feature to classify and detect obscene sounds.

## 4. REPEATED CURVE-LIKE SPECTRUM FEATURE

The repeated curve-like spectrum feature (RCSF) represents a repeated large temporal variation of frequency spectrum [14]. The large temporal variation of a frequency spectrum was modeled as a two-dimensional mel-frequency cepstrum coefficient (TDMFCC), which was originally used for automatic bird species classification to represent both the static (instantaneous cepstrum) and dynamic features (temporal variations) of bird sounds [15]. However, the syllable-based approach is not appropriate in our studies because of the nature of our data, which is collected in somewhat noisy environments, making it extremely difficult to automatically detect the syllable-segment. Therefore, we propose the RCSF feature as a modification of the original TDMFCC in order to make the method more effective. In addition, we added repeated curve-like properties found in general videos and noisy environments by extracting features in a long-range segment with a fixed length instead of a syllable-segment with variable length.

We defined the following terms in order to construct the RCSF feature vector in this paper. A *frame* is defined as a unit in which the sound property is not changed, that is, the frequency property is stationary. We used 32ms as a frame length, which has been adopted in most speech recognition systems. A *segment* is defined as a unit representing the temporal variation of the frequency spectrum, that is, it has curve-like spectrum property. We used 500ms as a segment length because most curve-like shapes of obscene sounds are about 500ms in length. A *clip* is defined as a unit that represents the repeated property of curve-like spectrum. An RCSF feature vector is constructed in a clip unit. The length of a clip is 10s because the repeated property of curve-like shapes is shown well at this length.

The procedure of constructing the RCSF feature vector is as follows. First, each audio clip is divided into several segments composed of a fixed number of frames. In this paper, a 500ms segment length and 32ms frame length with a 50% overlap are used. Second, MFCCs are computed in each frame, and the $q$-th MFCC of the $t$-th frame of $k$-th segment, $C_{k,t}(q)$, is calculated as (1), where $E_{k,t}(b)$ is the energy of the $b$-th mel-scaled bandpass filter of the $t$-th frame of the $k$-th segment, $q$ is the quefrency index, and $B$ is the number of mel-scale bandpass filters.

$$C_{k,t}(q) = \sum_{b=0}^{B-1} \log(E_{k,t}(b)) \cos(\frac{(2b+1)q\pi}{2B}) \tag{1}$$

Next, the curve-like spectrum feature of a segment can be calculated by applying a 1-dimension DCT to a sequence of $L$ successive MFCCs along the time axis within the fixed-length segment. Matrix $C_k(q,n)$ of the curve-like spectrum feature of the $k$-th segment can be calculated as (2), where $n$ is the index of the modulation frequency, $L$ is the number of frames within a fixed-length segment (32 frames with a 50% overlap within a 500ms segment), and $K$ is the number of segments in a clip.

$$C_k(q,n) = \sum_{t=0}^{L-1} C_{k,t}(q) \cos(\frac{(2t+1)n\pi}{2L}), \quad 0 \leq q < B, 0 \leq n < L, 0 \leq k < K \tag{2}$$

Equation (2) indicates that the first row of $C_k(q,n)$ represents the temporal variations of the short-time energy within a long-range $k$-th segment. The first column represents the average MFCC of all frames in the $k$-th segment. More generally, along the quefrency axis $q$ (i.e., column axis), the lower-order coefficients describe the spectral envelop, and the higher-order coefficients describe extra information such as pitch. Along the time axis $n$ (i.e. row axis), the lower-order coefficients represent the global variations of the MFCC and the higher-order coefficients represent the local variations of the MFCC. In this paper, various RCSF feature vectors configured by combining with various numbers of coefficient orders in the quefrency side (i.e., column axis) and time side (i.e., row axis) were tested in order to check which order of quefrency and time are more effective to classify obscene sounds. The feature vector of the $k$-th segment, $F_{seg\_k}$, is constructed with the chosen order of the quefrency ($B'$) side and time ($L'$) side as (3).

$$F_{seg\_k} = [C_k(0,0),...,C_k(0,L'-1), C_k(1,0),...,C_k(1,L'-1), C_k(2,0),...,C_k(2,L'-1),...,C_k(B'-1,0),...,C_k(B'-1,L'-1)]^T \tag{3}$$

Finally, the RCSF feature vector, $F_{RCSF}$, is constructed with mean and standard deviation values of all of the curve-like spectrum feature vectors of segment $F_{seg\_k}$, in order to apply the repeated properties as (4).

$$F_{RCSF} = [mean(F_{seg\_k}), std(F_{seg\_k})]^T, \quad 0 \leq k \leq K \tag{4}$$

## 5. CLASSIFICATION WITH SVM CLASSIFIER

A support vector machine (SVM) classifier is used in order to train RCSF feature vectors and classify whether an unknown input sample is obscene. Because there are very many different samples within obscene sounds even though the category of obscene sounds is limited to sexual moans and screams, it is important that the classifier have a generalized performance. The SVM classifier can also make the feature vectors trained so that the decision function maximizes the generalization ability [16]. SVM is a binary classifier that makes its decisions by constructing an optimal hyperplane that separates the two classes with the largest margin. It is based on the idea of structural risk minimization induction principle that aims at minimizing a bound on the generalization error, rather than minimizing the mean square error [17]-[20]. For the optimal hyperplane $\mathbf{w} \cdot \mathbf{x} + b = 0$, where $\mathbf{w} \in R^N$ and $b \in R$, the decision function of classifying a unknown point $\mathbf{x}$ is defined as (5), where $N_S$ is the support vector number, $\mathbf{x}_i$ is the support vector, $\alpha_i$ is the Lagrange multiplier and $m_i \in \{-1, 1\}$ describes which class $\mathbf{x}$ belongs to.

$$f(\mathbf{x}) = sign(\mathbf{w}\mathbf{x} + b) = sign(\sum_{i=1}^{N_S} \alpha_i m_i \mathbf{x}_i \cdot \mathbf{x}) \tag{5}$$

In most cases, searching suitable hyperplane in input space is too restrictive to be of practical use. The solution to this situation is mapping the input space into a higher dimensional feature space and searching the optimal hyperplane in this feature space. Let $\mathbf{z} = \varphi(\mathbf{x})$ denote the corresponding feature space vector with a mapping $\varphi$ from $R^N$ to a feature space Z. It is not necessary to know about $\varphi$. We just provide a function called kernel which uses the points in input space to compute the dot product in feature space Z that is (6).

$$\mathbf{z}_i \cdot \mathbf{z}_j = \varphi(\mathbf{x}_i) \cdot \varphi(\mathbf{x}_j) = K(\mathbf{x}_i, \mathbf{x}_j) \tag{6}$$

Finally, the decision function becomes (7).

$$f(\mathbf{x}) = sign(\sum_{i=1}^{N_S} \alpha_i m_i K(\mathbf{x}_i, \mathbf{x}) + b) \tag{7}$$

Typical kernel functions include linear kernel, polynomial, radial basis kernel, etc.

In this paper, the '*libsvm*' tools are used as an SVM classifier [21], and the radial basis kernel function (RBF) is used as a kernel function. The RBF kernel nonlinearly maps samples into a higher dimensional space, so unlike the linear kernel, it can handle a case in which the relation between class labels and attributes is nonlinear.

The procedures of training and classifying RCSF feature vectors extracted from the objectionable and non-objectionable audio clips are as follows. First, feature vectors for training and generating the learning model are extracted from the training audio clips and are scaled within the range of {-1, 1}. It is very important to scale the feature vectors before training and classifying them with the SVM classifier. The main advantage is to avoid allowing the attributes in greater numeric ranges to dominate those in smaller numeric ranges. Another advantage is to avoid numerical difficulties during the calculation [22]. After scaling the feature vectors, their optimized parameters are chosen through cross validation for training. We use 5-fold cross validations in this paper. The optimized parameters make the classifier able to more accurately predict unknown data (i.e., testing data). Finally, the learning model is generated using training feature vectors with optimized parameters. This model is referred to when the unknown data is predicted. Feature vectors of test audio clips are constructed in the same way as in the training audio clips, and are scaled within the same range. In addition, the classification is performed by predicting the class of feature vectors extracted from test audio clips based on the previously generated learning model. If an audio clip of an obscene sound is predicted as an obscene sound, or an audio clip of a non-obscene sound is predicted as a non-obscene sound, the clip is classified correctly. Otherwise, it is an incorrect classification.

## 6. CONSTRUCTION OF DATASET

### 6.1. Dataset of audio clips

To evaluate the RCSF feature and its learning model, we constructed a dataset of audio clips that have two classes, an obscene class and a non-obscene class, including a total of 6,269 audio clips. All of the audio clips are 10s long and are digitized at 16 bits per sample with a 16,000Hz sampling rate in a mono channel. The scale of the dataset in this paper is larger than that of the early tests of [13], [14] and balances the number of clips between the two classes so that classification performance is not biased to the one class that has the larger number of clips.

Audio clips of an obscene class are constructed according to the 6 categories mentioned in section 3. There are 3,052 audio clips of the obscene class (2,040 for training and 1,012 for testing) collected from three types of videos: adult videos (85), adult broadcasts (87), and hidden or self-recorded videos (39). The numbers in the parenthesis indicate the number of video clips for each type. Since the contents of self-recorded and hidden-recorded are very similar, and it is very difficult to distinguish them, we categorized the two types of content into the same group. The detailed configuration of audio clips of obscene class is shown in Table 1. All clips of an obscene class are collected manually from a variety of X-rated videos so that they include various types of sexual moans and screams. The distribution of clips as shown in Table 1 shows that obscene sounds consist mainly of female-based sounds at more than 90%. It is also important to configure the audio clips of a non-obscene class in order to cover almost all of the non-obscene area. Thus, we try to collect audio clips of the non-obscene class from most genres of video clips and audio tracks that are distributed in real fields. We divided videos of the non-obscene class into eight genres: soap operas and movies (87), current events and culture (27), sports (17), children (23), entertainment (43), news (10), pop music (196), and instrumental music (84). There were 3,217 audio clips of the non-obscene class collected from the above genres (2,089 for training and 1,128 for testing). Although the number of video clips of each genre is different, we collected a similar number of audio clips for each genre. The detailed configurations of audio clips of non-obscene class are shown in Table 2.

Table 1. Configuration of audio clips of obscene class

| Categories of clips | Clips for training | | Clips for testing | |
| --- | --- | --- | --- | --- |
| | # | Rate(%) | # | Rate(%) |
| BOSM | 424 | 21 | 206 | 20 |
| BOSS | 285 | 14 | 178 | 17 |
| FMSM | 650 | 32 | 311 | 31 |
| FMSS | 523 | 26 | 219 | 22 |
| MASM | 106 | 5 | 68 | 7 |
| MASS | 52 | 2 | 30 | 3 |
| Total | 2,040 | 100 | 1,012 | 100 |

Table 2. Configuration of audio clips of non-obscene class

| Genres of clips | Clips for training | | Clips for testing | |
| --- | --- | --- | --- | --- |
| | # | Rate(%) | # | Rate(%) |
| Soap operas & Movies | 300 | 14 | 215 | 19 |
| Current events & Cultures | 252 | 12 | 117 | 10 |
| Sports | 253 | 12 | 120 | 11 |
| Children | 250 | 12 | 132 | 12 |
| Entertainments | 264 | 13 | 126 | 11 |
| News | 270 | 13 | 100 | 9 |
| Pop music | 250 | 12 | 196 | 17 |
| Instrumental music | 250 | 12 | 122 | 11 |
| Total | 2,089 | 100 | 1,128 | 100 |

Because all of the clips, especially for the obscene class, were collected from real videos, they include some environmental and recording noises, for example, fricative from bodies, background music, a loud laugh, creak of beds, talking during sexual activities, and so on. Recording noises are mostly included in the clips of the obscene class since many X-rated videos are made by amateurs in poor conditions in comparison with general videos. Therefore, the test results from the above datasets can be treated similarly as if the tests were performed under a noisy environment. However, because the noise strength including the datasets cannot be estimated and the noise is included at random, we performed the classification tests using noisy data generated by adding Gaussian white noise with a 5dB signal-to-noise ratio (SNR) into the test audio clips described in Tables 1 and 2. Gaussian white noise is a good approximation of many real world situations and generates mathematically traceable models. We can guess the amount of noise in a signal with a specific SNR (dB) value. Within noisy clips with a 5dB SNR, the average amplitude of the main signal is only 1.77 times larger than that of the noise signal, and these noisy clips are significantly harsh enough to hear.

### 6.2. Dataset of videos

In order to evaluate the classification performance of videos with the learning model generated from audio clips based on the RCSF feature, we construct the dataset of videos that are not used to construct the dataset of audio clips. A classification test of the video level is needed in order to check that the learning model based on the RCSF feature can be appropriate to decide whether an arbitrary video is X-rated or not. Each video file is gathered from some web-hard sites and peer-to-peer programs. The dataset of general and X-rated videos are configured as shown in Tables 3 and 4, respectively. Each genre of X-rated videos includes commercial videos made by professionals, as well as personal videos made by amateurs, such as self or hidden recorded videos.

Table 3. Configuration of general videos.

| Genres of videos | The number | Average running time(min.) |
| --- | --- | --- |
| Korean soap operas | 65 | 57 |
| Japanese soap operas | 65 | 49 |
| Western soap operas | 62 | 37 |
| Korean movies | 55 | 105 |
| Western movies | 62 | 89 |
| Sports | 62 | 42 |
| Music Videos | 49 | 4 |
| Documentaries | 60 | 42 |
| Entertainments | 60 | 64 |
| News | 60 | 36 |
| Total | 600 | |

Table 4. Configuration of X-rated videos.

| Genres of videos | The number | Average running time(min.) |
|---|---|---|
| Korean adult videos | 260 | 40 |
| Japanese adult videos | 215 | 75 |
| Western adults videos | 125 | 58 |
| Total | 600 | |

## 7. EXPERIMENTAL RESULTS

### 7.1. Experimental results for audio clips

In order to estimate the performance of the proposed RCSF feature, we compare the performance of RCSF feature with a variety of well-known spectral features [19], [22]-[25], which are MFCC; MFCC with delta coefficient (MFCCD); MFCC with delta and double delta coefficient (MFCCDD); spectral low-level features (LLF_S) composed of spectral bandwidth, spectral centroid, spectral flatness, spectral flux, and spectral roll-off set at 85%; and spectral and energy low-level features (LLF_ES) composed of the total energy and eight sub-band energies of frequency in addition to the LLF_S features. In order to investigate how the order affects the classification performance, the MFCC, MFCCD, and MFCCDD features were tested for 9 order types, from 7 to 23 at intervals of 2, and the RCSF features were tested for 72 combinations of 9 quefrency order types, from 7 to 23, and 8 temporal order types, from 5 to 19, both at intervals of 2. Each feature vector was constructed from a 10s clip unit of the dataset, which were configured in 6.1, using the mean and standard deviation values of all of the feature vectors that were extracted from each segment of a 500ms length within a clip.

The performance was estimated in terms of F1-score, precision, and recall, which were computed using (8), (9), and (10), respectively, where $N_{tp}$ is the number of true positives, $N_{tn}$ is the number of true negatives, $N_{fp}$ is the number of false positives, and $N_{fn}$ is the number of false negatives. Obscene audio clip is used as positive data in this paper. Therefore, a true positive means that a clip classified as obscene class is an obscene audio clip, and a false positive means that a clip classified as obscene class is a non-obscene audio clip.

$$F1\text{-}score(\%) = \frac{2 \cdot Precision \cdot Recall}{Precision + Recall} \times 100 \quad (8)$$

$$Precision(\%) = \frac{N_{tp}}{N_{tp} + N_{tn}} \times 100 \quad (9)$$

$$Recall(\%) = \frac{N_{tp}}{N_{tp} + N_{fn}} \times 100 \quad (10)$$

The overall performance of each feature is represented by the F1-score, as this is a balanced performance of precision and recall. A score of 1 is the best F1 score, while a 0 indicates the worst F1 score. Precision represents the ratio of obscene audio clips classified correctly among all clips classified as obscene class, while recall represents the ratio of obscene audio clips classified correctly among all clips tested as obscene class.

Figure 3 shows the best classification performance of the RCSF and other well-known features on an F1-score basis at the original dataset. The RCSF has its best performance at a quefrency order of 23 and temporal order of 15. MFCC, MFCCD, and MFCCDD have their best performance at a quefrency order of 23, 23, and 21, respectively. Figure 4 shows the classification performance of the same features as shown in Fig. 3 with noisy data of 5dB SNR.

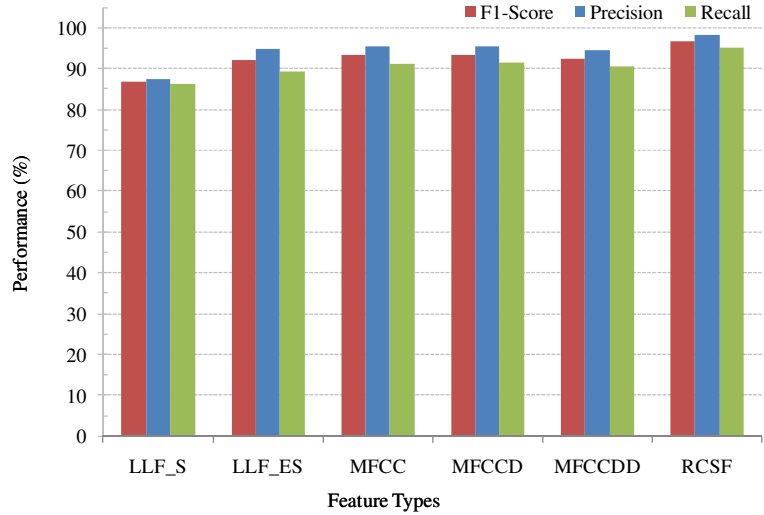

Figure 3. Classification performance of various features at the original dataset

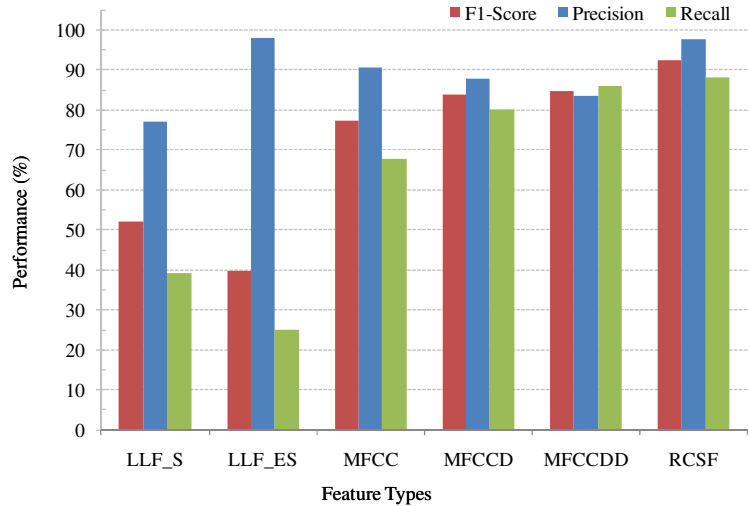

Figure 4. Classification performance of various features at the noisy dataset of 5dB SNR

The proposed RCSF feature outperforms LLF, MFCC, MFCCD, and MFCCDD features by 9.73%, 4.53%, 3.35%, 3.25%, and 4.16% in the original dataset, and 40.55%, 52.86%, 15.11%, 8.88%, and 7.78% in the noisy dataset, respectively. Particularly in a noisy dataset, the difference in classification performance between the RCSF and other features is larger. Figure 4 shows that the RCSF feature is also robust in the noisy dataset as it worsens by only 4.09% compared with the F1-score in the original dataset, and still maintains 92.55% of the F1-score even though the precision and recall worsen by 0.59% and 7.15%, respectively. This result shows that the RCSF is a stable and proper feature for classifying obscene sounds. The temporal variation of spectrum represented by RCSF is also very important information for classifying obscene sounds.

Each performance of LLF_S, LLF_ES, MFCC, MFCCD and MFCCDD is not so bad for original dataset, but becomes unstable in the noisy dataset as shown in Fig 4. In particular, the LLF_S and LLF_ES feature have very poor performances. Recall is very low and then F1-score is also low even though precision is high, that is, the probability that a clip determined as obscene class is correctly classified is very high, but the probability for an obscene audio clip to be classified as obscene class is very low. If LLF_S and LLF_ES are applied to filtering X-rated videos, many X-rated videos fail to be filtered. There is little difference among the performances of MFCC, MFCCD, and MFCCDD in the original dataset. However, in the noisy dataset, MFCCD and MFCCDD have more stable and better performances than MFCC because MFCCD and MFCCDD include the delta values that represent the temporal variation of spectrum between successive frames, and these delta values are not included in MFCC. This result shows that one important piece of information that can classify an obscene sound from other sounds is an inherent temporal variation of spectrum, that is, a repeated curve-like pattern, and the proposed RCSF feature is more suitable to classify an obscene sound than other features.

Classification errors of the RCSF feature occurred in the following cases. The cases of classifying non-obscene audio clips as obscene class often occurred in the genres of soap operas, movies, and talk shows. In soap operas and movies, exaggerated intonations and accents sometimes have similar spectrum patterns with obscene sounds. For talk shows, some of the test clips included signals with high pitches such as the responses and laughter of audiences, and thus these clips were often misjudged. In particular, female laughter is a signal that creates a lot of false classification. The cases of classifying obscene audio clips as non-obscene class often occurred for signals from weak sexual moans. Because the magnitude of signals for most weak sexual moans is very small, the temporal variation of spectrum is too small to distinguish well. In particular, some signals of male sexual moans are often misjudged because they have a relatively low pitch and small temporal variation of spectrum.

Table 5 lists the performance of the top 5 feature types and the bottom 5 ones among all of 72 RCSF feature types. And tables 6, 7, and 8 list the performances of all types of MFCC, MFCCD, and MFCCDD in descending order of F1-score. *Orig.* and *5dB* in each table mean the performances at the original dataset and at the noisy dataset in 5dB SNR respectively. RCSF features have more stable performance than MFCC, MFCCD, and MFCCDD features. In fact, the average F1-score of all of 72 RCSF feature types is 95.38% with 0.69 of standard deviation at the original dataset and 89.88% with 2.97 of standard deviation at the noisy dataset. Especially RCSF feature outperforms the MFCC, MFCCD, and MFCCDD features at the noisy dataset. These results verify again that RCSF is a proper feature for classifying obscene sounds. As the quefrency order of RCSF feature becomes higher, the performance is very slightly improved. But the size of temporal order itself does not affect to the performance. In other words, the important thing is the temporal property of frequency itself not the size of temporal order.

Table 5. Performance of RCSF feature types with combinations of quefrency order(Q) and temporal order(T).

| Q | T | F1-score(%) | | Precision(%) | | Recall(%) | |
|---|---|---|---|---|---|---|---|
|   |   | Orig. | 5dB | Orig. | 5dB | Orig. | 5dB |
| 23 | 15 | 96.64 | 92.55 | 98.17 | 97.58 | 95.16 | 88.01 |
| 23 | 13 | 96.60 | 92.38 | 97.87 | 97.57 | 95.36 | 87.71 |
| 23 | 11 | 96.55 | 92.17 | 97.77 | 97.35 | 95.36 | 87.51 |
| 23 | 9 | 96.54 | 91.99 | 97.97 | 97.56 | 95.16 | 87.02 |
| 23 | 17 | 96.49 | 93.13 | 97.96 | 96.49 | 95.06 | 89.99 |
| … | … | … | … | … | … | … | … |
| 9 | 5 | 94.19 | 84.08 | 97.16 | 98.15 | 91.40 | 73.54 |
| 7 | 5 | 94.10 | 87.87 | 96.96 | 97.35 | 91.40 | 80.08 |

| 7 | 19 | 94.10 | 90.36 | 96.09 | 96.51 | 92.19 | 84.94 |
|---|---|---|---|---|---|---|---|
| 7 | 15 | 93.96 | 90.36 | 96.66 | 96.51 | 91.40 | 84.94 |
| 7 | 17 | 93.92 | 89.68 | 96.36 | 97.33 | 91.60 | 83.15 |
| Mean | | 95.38 | 89.88 | 97.12 | 97.32 | 93.72 | 83.70 |
| Std | | 0.69 | 2.97 | 0.51 | 1.06 | 1.28 | 5.38 |

Table 6. Performance of MFCC feature types with various quefrency order(Q).

| Q | F1-score(%) | | Precision(%) | | Recall(%) | |
|---|---|---|---|---|---|---|
| | Orig. | 5dB | Orig. | 5dB | Orig. | 5dB |
| 23 | 93.29 | 77.44 | 95.36 | 90.46 | 91.30 | 67.69 |
| 21 | 92.73 | 77.59 | 94.74 | 90.16 | 90.81 | 68.09 |
| 19 | 92.44 | 79.54 | 94.99 | 87.60 | 90.02 | 72.84 |
| 17 | 91.61 | 79.77 | 94.35 | 85.67 | 89.03 | 74.63 |
| 15 | 91.27 | 77.19 | 94.40 | 83.93 | 88.34 | 71.46 |
| 13 | 90.02 | 78.16 | 94.46 | 78.12 | 85.97 | 78.20 |
| 11 | 88.68 | 77.51 | 93.92 | 77.63 | 83.99 | 77.40 |
| 9 | 87.55 | 76.62 | 93.89 | 74.74 | 82.02 | 78.59 |
| 7 | 84.52 | 74.68 | 92.69 | 71.47 | 77.67 | 78.20 |
| Mean | 90.23 | 77.61 | 94.31 | 82.20 | 86.57 | 74.12 |
| Std | 2.87 | 1.52 | 0.77 | 6.93 | 4.59 | 4.34 |

Table 7. Performance of MFCCD feature types with various quefrency order(Q).

| Q | F1-score(%) | | Precision(%) | | Recall(%) | |
|---|---|---|---|---|---|---|
| | Orig. | 5dB | Orig. | 5dB | Orig. | 5dB |
| 23 | 93.39 | 83.67 | 95.46 | 87.72 | 91.40 | 79.98 |
| 21 | 93.27 | 83.09 | 95.64 | 87.76 | 91.01 | 78.89 |
| 19 | 92.40 | 81.54 | 94.90 | 83.72 | 90.02 | 79.48 |
| 17 | 92.08 | 82.39 | 94.68 | 80.90 | 89.62 | 83.94 |
| 15 | 91.54 | 80.12 | 94.43 | 80.56 | 88.83 | 79.68 |
| 13 | 90.17 | 80.44 | 93.06 | 74.79 | 87.45 | 87.02 |
| 11 | 89.37 | 76.33 | 93.05 | 76.75 | 85.97 | 75.92 |
| 9 | 87.79 | 78.27 | 93.03 | 72.57 | 83.10 | 84.94 |
| 7 | 85.55 | 75.08 | 91.74 | 70.83 | 80.14 | 79.88 |
| Mean | 90.62 | 80.10 | 94.00 | 79.51 | 87.50 | 81.08 |
| Std | 2.65 | 3.00 | 1.33 | 6.22 | 3.81 | 3.48 |

Table 8. Performance of MFCCDD feature types with various quefrency order(Q).

| Q | F1-score(%) | | Precision(%) | | Recall(%) | |
|---|---|---|---|---|---|---|
| | Orig. | 5dB | Orig. | 5dB | Orig. | 5dB |
| 21 | 92.48 | 84.77 | 94.53 | 83.54 | 90.51 | 86.03 |
| 23 | 92.41 | 84.50 | 94.71 | 86.11 | 90.22 | 82.95 |
| 19 | 91.75 | 83.66 | 93.44 | 82.34 | 90.12 | 85.03 |
| 17 | 91.24 | 83.21 | 93.56 | 81.29 | 89.03 | 85.23 |
| 15 | 90.95 | 81.27 | 94.27 | 83.79 | 87.85 | 78.89 |
| 13 | 89.53 | 80.11 | 93.16 | 74.51 | 86.17 | 86.62 |
| 11 | 88.97 | 78.53 | 93.00 | 73.10 | 85.28 | 84.84 |
| 9 | 87.30 | 77.60 | 91.82 | 71.78 | 83.20 | 84.44 |
| 7 | 86.95 | 78.07 | 91.77 | 68.43 | 82.61 | 90.88 |
| Mean | 90.17 | 81.30 | 93.36 | 78.32 | 87.22 | 84.99 |
| Std | 2.09 | 2.84 | 1.07 | 6.38 | 3.05 | 3.17 |

## 7.2. Experimental results for videos

For classifying videos using the RCSF feature and its model, we use a *harmful rate* that is defined as the ratio of obscene audio clips among the total audio clips within a particular video, which is computed using (11).

$$Harmful\ rate(\%) = \frac{\#\ of\ obscene\ audio\ clips}{\#\ of\ all\ audio\ clips} \times 100 \qquad (11)$$

We estimate the harmful rate for 1,200 videos configured as shown in Tables 3 and 4 by means of three types of RCSF features that are constructed using a quefrency and temporal order pair: (9, 7), (15, 7), (21, 7). In order to investigate the difference between the estimation results of each RCSF feature type, we used three types of quefrency orders: 9 for lower order, 15 for middle order, and 21 for higher order. Figure 5 shows the histogram of the harmful rate for videos analyzed by means of each RCSF feature type. *GEN* and *X* indicate a general video and a X-rated video respectively. *RCSF* means that the RCSF feature is used as an audio feature, and *Q* and *T* indicate the quefrency and temporal order, respectively. For example, *GEN_RCSF_Q(21)_T(7)* means that a general video is analysed by means of the RCSF feature constructed with a quefrency order of 21 and temporal order of 7. In Fig. 5, the three types of RCSF features have similar results for both general and X-rated videos. More than 82% of the general videos have a harmful rate of less than 10%, and more than 97% of general videos have a harmful rate of less than 20%. Unlike the expectation of a particular minimum harmful rate, all X-rated videos have a harmful rate between 0% and 100%.

However, the harmful rate of X-rated videos is more distributed within the range of 40% to 90%. Most of the X-rated videos that have a low harmful rate do not have an obscene sound. For example, in the case of personal videos made by amateurs under poor environments, such as self or hidden recorded videos, the audio signal is very weak or does not exist. While obscene sound exists, the amount of environmental and recording noises is too large to detect an obscene sound. Some of the professional videos include mostly background music and sound effects as they are focused on the body exposure of actors or actresses rather than on sexual activities. Also, several videos have been edited in order to focus mainly on sexual conversations. From the histogram in Fig. 5, we adopted a 20% harmful rate as the threshold for classifying X-rated videos.

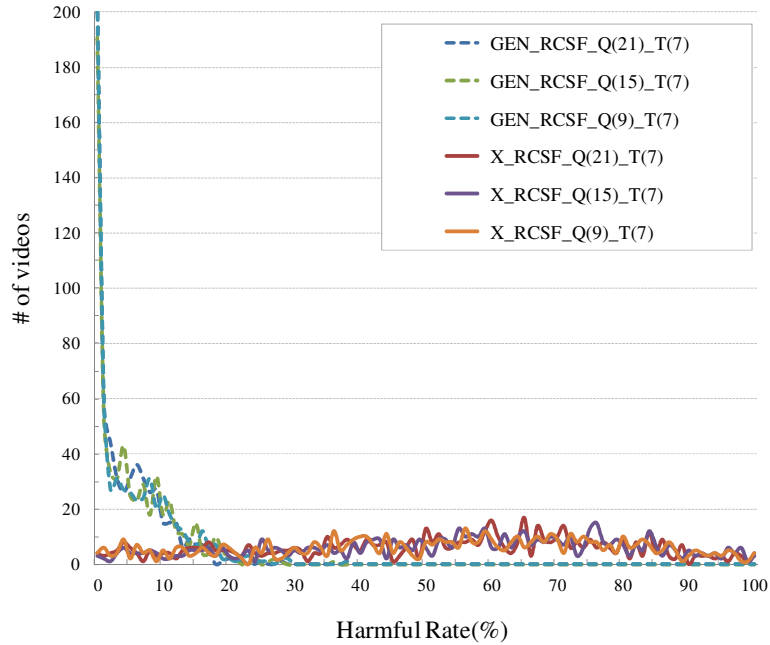

Figure 5. Histogram of the harmful rate for 600 X-rated videos and 600 general videos

Table 9 shows the classification performance of three types of RCSF features with a 20% harmful rate as the threshold for 1,200 videos. The three types of RCSF features have a similar performance, and the precision is particularly high, while the recall is relatively low. In other words, while the possibility of detecting an X-rated video is relatively low, the probability that a video classified as X-rated was done so correctly is quite high. Tables 10 and 11 show the classification error rates of X-rated and general videos, respectively, for each genre. The numbers in the parenthesis indicate the number of videos.

Table 9. Classification performance for videos at 20% of harmful rate as threshold

| Feature Type | F1-score(%) | Precision(%) | Recall(%) |
|---|---|---|---|
| Q(9)_T(7) | 90 | 97 | 85 |
| Q(15)_T(7) | 92 | 98 | 87 |
| Q(21)_T(7) | 91 | 99 | 84 |

Table 10. Classification error rate of X-rated videos

| Genres | The number | Classification error rate(%) | | |
|---|---|---|---|---|
| | | Q(9)_T(7) | Q(15)_T(7) | Q(21)_T(7) |
| Korean adult videos | 260 | 21.9(57) | 18.4(48) | 22.3(58) |
| Japanese adult videos | 215 | 6.0(13) | 4.6(10) | 7.4(16) |
| Western adults videos | 125 | 16.8(21) | 15.2(19) | 16.0(20) |
| Total | 600 | 15.1(91) | 12.8(77) | 15.6(94) |

Table 11. Classification error rate of general videos

| Genres | The number | Classification error rate(%) | | |
|---|---|---|---|---|
| | | Q(9)_T(7) | Q(15)_T(7) | Q(21)_T(7) |
| Korean soap operas | 65 | 1.5(1) | 0.0(0) | 0.0(0) |
| Japanese soap operas | 65 | 0.0(0) | 0.0(0) | 0.0(0) |
| Western soap operas | 62 | 0.0(0) | 0.0(0) | 0.0(0) |
| Korean movies | 55 | 12.7(7) | 10.9(6) | 1.8(1) |
| Western movies | 62 | 9.7(6) | 11.3(7) | 6.5(4) |
| Sports | 62 | 0.0(0) | 0.0(0) | 0.0(0) |
| Music Videos | 49 | 4.1(2) | 0.0(0) | 2.0(1) |
| Documentaries | 60 | 0.0(0) | 0.0(0) | 0.0(0) |
| Entertainments | 60 | 0.0(0) | 0.0(0) | 0.0(0) |
| News | 60 | 0.0(0) | 0.0(0) | 0.0(0) |
| Total | 600 | 2.7(16) | 2.2(13) | 1.0(6) |

The reasons for a higher error rate of X-rated videos are mentioned above. There are more X-rated videos that have few obscene sounds or very weak audio signals than we expected. Also, hidden and/or self-recorded videos that have low-quality audio as well as poor visual information were included in Korean and Western adult videos, but not Japanese adult videos. Therefore, the classification error rate of Japanese adult videos is lower than the others. For general videos, there are few errors to classify them. As expected, there are a few errors for soap operas and movies as exaggerated intonations and accents appeared mostly in these genres. Also, laughter, especially women's laughter, appeared in these same genres.

From the results, X-rated videos can be classified by means of the RCSF feature with a low error rate, and can be classified more effectively and precisely if visual features are adopted to classify X-rated videos that have few obscene sounds or very weak audio signals.

## 8. CONCLUSION

In this paper, we propose the repeated curve-like spectrum feature as an audio feature for classifying obscene sounds and X-rated videos. Obscene sounds indicate an audio signal generated from sexual moans and screams in various sexual scenes. For a reasonable evaluation of the RCSF feature, we constructed two types of datasets using two classes, obscene and non-obscene content. The first type of dataset is composed of audio clips. Audio clips of the obscene class were composed with six categories defined from analyzing various samples of obscene sounds, and those of the non-obscene class were composed to represent general audio signals. The second type of dataset is composed of various genres of videos. The classification for audio clips was performed using a support vector machine, and the classification of videos was performed using a harmful rate defined as the ratio of obscene clips to the total clips within a particular video. The repeated curve-like spectrum feature has an F1-score, precision, and recall rate of about 96.64%, 98.18%, and 95.16%, respectively, in the original dataset, and 92.55%, 97.58%, and 88.01% in a noisy dataset with a 5dB SNR. Also, the proposed feature outperformed other well-known audio features, low-level perceptual features, and mel-frequency cepstrum coefficient families, by 3.25% to 9.73% in the original dataset, and 7.78% to 52.86% in the noisy dataset. This result shows that the repeated curve-like spectrum feature is a proper feature to classify obscene sounds and has a stable performance in noisy environments. For classifying videos based on the proposed feature, the classification performance has more than a 90% F1-score, 97% precision, and an 84% recall rate. Using the measured performance, the RCSF feature can be used to classify X-rated videos as well as obscene sounds. In addition, X-rated videos can be classified effectively with only audio features.

## ACKNOWLEDGEMENTS

This research was supported by the KCC(Korea Communications Commission), Korea, under the R&D program supervised by the KCA(Korea Communications Agency)"(KCA-2011-09914-06003).

## REFERENCES

[1] P. M. Kamde, and S. P. Algur, (2011) "A Survey on Web Multimedia Mining", *The International Journal of Multimedia & Its Applications(IJMA)*, Vol.3, No.3, pp.72-84.

[2] J. Coopersmith, (2000) "Pornography, videotape, and the internet", IEEE *Technology and Society Magazine*, pp.27-34.

[3] J. Z. Wang, J. Li, G. Wiederhold, and O. Firschein, (1998) "System for screening objectionable images", *Journal of Images and Computer Communications*, Vol.21, pp.1355-1360.

[4] H.A. Rowley, Y. Jing, and S. Baluja, (2006) "Large Scale Image-based Adult Content Filtering", *Proc. International Conf. on Computer Vision Theory and Applications(VISAPP)*, pp.290-296.

[5] Google Sage Search, http://www.google.com. Accessed 15 March 2010.

[6] A. Bosson, G. C. Cawley, Y. Chan, and R. Harvey, (2002) "Non-retrieval: blocking pornographic images", *Proc. ACM International Conf. on Image and Video Retrieval*, pp.50-60.

[7] J. S. Lee, Y. M. Kuo, P. C. Chung, and E. L. Chen, (2007) "Naked image detection based on adaptive and extensible skin color model", *Journal of Pattern Recognition*, Vol.40, pp.2261-2270.

[8] C. Jansohn C, A. Ulges, and T. M. Breuel, (2009) "Detecting Pornographic Video Content by Combining Image Features with Motion Information", *Proc. ACM International Conference on Multimedia*.

[9] C. Y. Kim, O. J. Kwon, W. G. Kim, and S. R. Choi, (2008) "Automatic system for filtering obscene video", *Proc. 10th International Conference on Advanced Communication Technology(ICACT2008)*. pp.1435-1438.

[10] Z. Liu, Y. Wang, and T. Chen, (1998) "Audio Feature extraction and analysis for scene segmentation and classification", *Journal of VLSI Signal Processing System*, vol.20, no.1/2, pp.61-79.

[11] N. Rea, G. Lacey, C. Lambe and R. Dahyot, (2006) "Multimodal Periodicity Analysis for Illicit Content Detection in Videos", *Proc. IET Conference on Visual Media Production(CVMP2006)*, pp.106-114.

[12] D. U. Cho, (2004) "Blocking of Internet Harmful Pronographic Sites by Contents-based Method", *The journal of Korea Information and Communiation Society(KICS)*, vol.29, No.6B, pp.554-562.

[13] J. D. Lim, S. W. Han, B. C. Choi, and B. H. Chung, (2009) "A study of classifying the objectionable contents by using the MFCCs-based audio feature set", *Proc. 10th International Conf. on Computers, Communications, and Systems (ICCCS'09)*, pp.255-258.

[14] J. D. Lim, B. C. Choi, S. W. Han, B. H. Chung and C. H. Lee, (2011) "Classification and Detection of Objectionable Sounds Using Repeated Curve-like Spectrum Feature", *Proc. International Conf. on Information Science and Applications(ICISA2011)*, On-line Proceedings.

[15] C. H. Lee, C. C. Han, C. C. Chuang, (2008) "Automatic Classification of Bird Species From Their Sounds Using Two-Dimensional Cepstral Coefficients", IEEE *Transactions on Audio, Speech, and Language Processing*, vol.16, no.8, pp.1541-1550.

[16] A. Shigeo, (2005) *Support Vector Machines for Pattern Classification*, Springer-Verlag London.

[17] C. Cortes and V. Vapnik, (1995) "Support Vector Networks", *Machine Learning*, vol.20, pp.273-297.

[18] V. N. Vapnik, (1998) *Statistical Learning Theory*, New York: Wiley.

[19] J. C. Wang, J. F. Wang, C. B. Lin, K. T. Jian, W. H. Kuok, (2006) "Content-based Audio Classification Using Support Vector Machines and Independent Component Analysis", *Proc. 18th International Conf. on Pattern Recognition(ICPR06)*, Vol.4, pp.157-160.

[20] C. S. Rao, S. S. Kumar, and B. C. Mohan, (2010) "Content Based Image Retrieval Using Exact Legendre Moments and Support Vector Machine", *The International Journal of Multimedia & Its Applications(IJMA)*, Vol.2, No.2, pp.69-79.

[21] C. W. Hsu, C. C. Chang, and C. J. Lin, (2009) "A Practical Guide to Support Vector Classification", http://www.csie.ntu.edu.tw/~cjlin/libsvm/, Accessed 30 Oct 2009.

[22] L. Lu, H. J. Zhang, S. Z. Li, (2003) "Content-based Audio Classification and Segmentation by Using Support Vector Machines", ACM *Multimedia Systems Journal,* Vol.8. no.6, pp.482-492.

[23] G. Guo, and S. Z. Li, (2003) "Content-based Audio Classification and Retrieval by Support Vector Machine", IEEE *transactions on Neural Networks*, vol.14, no.1, pp.209-215.

[24] D. Brezeale, and D. J. Cook, (2008) "Automatic Video Classification: A Survey of the Literature", IEEE *Transactions on Systems, Man, and Cybernetics_Part C: Applications and reviews*, vol.38, no.3, pp.416-430.

[25] L. Lu, H. J. Zhang, (2002) "Content analysis for audio classification and segmentation", IEEE *Transactions on speech and audio processing*, Vol.10, pp.504-516.

**Authors**

**JaeDeok Lim** had received my BS and MS degrees in Electronic Engineering from Kyungbook National University in 1999 and 2001 respectively. He is currently a member of engineering staff at knowledge-based Information Security Research Division in ETRI(Electronics and Telecommunications Research Institute). His research areas include signal & image processing, content classification, access control.
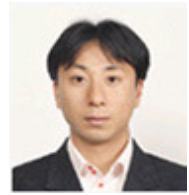

**Byeongcheol Choi** received the B.S. and M.S. degrees in Computer Science from University of Seoul, Korea in 1999 and 2001 respectively. From 2001, he has been a senior researcher at Electronics and Telecommunications Research Institute (ETRI), where he has developed the security schemes and contents filtering systems. His research interests in image/video processing, pattern recognition, and information security.
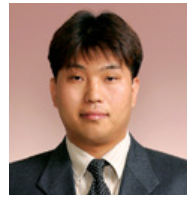

**Seung-Wan Han** received the B.S., M.S. and Ph.D. degree in Computer Science from Chonnam National University, Gwangju, Korea, in 1994, 1996 and 2001, respectively. Currently, he is a senior member of engineering staff at Electronics and Telecommunications Research Institute (ETRI) in Korea. His research interests include pattern classification, information security, and theory of computation.
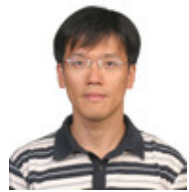

**Cheol-Hoon Lee** received the B.S. degree in electronics engineering from Seoul National University, Seoul, Korea, in 1983 and the M.S. and Ph.D. degrees in computer engineering from KAIST, Daejeon, Korea, in 1988 and 1992, respectively. From 1983 to 1994, he worked for Samsung Electronics Company in Seoul, Korea, as a researcher. From 1994 to 1995 and from 2004 to 2005, he was with the University of Michigan, Ann Arbor, as a research scientist at the Real-time Computing Laboratory. Since 1995, he has been a professor in the Department of Computer Engineering, Chungnam National University, Daejeon, Korea. His research interests include parallel processing, operating system, real-time system, and fault tolerant computing.
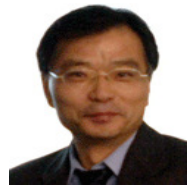